# Deep learning-based color holographic microscopy

*Tairan Liu[1,2,3†], Zhensong Wei[1†], Yair Rivenson[1,2,3†,*], Kevin de Haan[1,2,3], Yibo Zhang[1,2,3], Yichen Wu[1,2,3], and Aydogan Ozcan[1,2,3,4,*]*

[†] Equally contributing authors

[*] Corresponding authors: rivensonyair@ucla.edu ; ozcan@ucla.edu

[1]Electrical and Computer Engineering Department, University of California, Los Angeles, CA, 90095, USA.
[2]Bioengineering Department, University of California, Los Angeles, CA, 90095, USA.
[3]California NanoSystems Institute (CNSI), University of California, Los Angeles, CA, 90095, USA.
[4]Department of Surgery, David Geffen School of Medicine, University of California, Los Angeles, CA, 90095, USA.

**Abstract**

We report a framework based on a generative adversarial network (GAN) that performs high-fidelity color image reconstruction using a single hologram of a sample that is illuminated simultaneously by light at three different wavelengths. The trained network learns to eliminate missing-phase-related artifacts, and generates an accurate color transformation for the reconstructed image. Our framework is experimentally demonstrated using lung and prostate tissue sections that are labeled with different histological stains. This framework is envisaged to be applicable to point-of-care histopathology, and presents a significant improvement in the throughput of coherent microscopy systems given that only a single hologram of the specimen is required for accurate color imaging.

# 1. INTRODUCTION

Histological staining of fixed, thin tissue sections mounted on glass slides is one of the fundamental steps required for the diagnoses of various medical conditions. Histological stains are used to highlight the constituent tissue parts by enhancing the colorimetric contrast of cells and subcellular components for microscopic inspection. Thus, an accurate color representation of the stained pathology slide is an important prerequisite to make reliable and consistent diagnoses [1–4]. Unlike bright-field microscopy, a common method used to obtain color information from a sample using a coherent imaging system requires the acquisition of at least three holograms at the red, green, and blue parts of the spectrum, thus forming the red–green–blue (RGB) color channels that are used to reconstruct composite color images. Such colorization methods used in coherent imaging systems suffer from color inaccuracies [5–7] and may be considered unacceptable for histopathology and diagnostic applications.

To achieve increased color accuracy using coherent imaging systems, a computational hyperspectral imaging approach can be used [8]. However, such systems typically require engineered illumination, such as a tunable laser to efficiently sample the visible band. Previous contributions have demonstrated successful reduction in the number of required sampling locations for the visible band to generate accurate color images. For example, Peercy *et al*. demonstrated a wavelength selection method using Gaussian quadrature or Riemann summation for reconstructing color images of a sample imaged in reflection mode holography [5], whereby it was suggested that a minimum of four wavelengths were required to generate accurate color images of natural objects. Later, Ito *et al*. demonstrated a Wiener estimation-based method to quantify the spectral reflectance distribution of the object at four fixed wavelengths that achieved an increased color accuracy for natural objects [9]. Recently, Zhang *et al*. presented an absorbance spectrum estimation method based on minimum mean-square-error estimation, specifically crafted to create accurate color images of pathology slides with in-line holography [7]. Because the color distribution within a stained histopathology slide is constrained by the colorimetric dye combination that is used, this method successfully reduced the required number of wavelengths to three, while it still preserved accurate color representation. However, owing to the distortions introduced by twin image artifacts and the limited resolution of unit magnification on-chip holography systems, multiheight phase recovery [10–15] and pixel super-resolution (PSR) techniques [16–26] were implemented to achieve acceptable image quality.

Herein, we present a deep learning-based accurate color holographic microscopy method (**Figure 1**). In comparison to the traditional hyperspectral imaging approaches used in coherent imaging systems, the proposed deep neural-network-based color microscopy method significantly simplifies the data acquisition procedures, the associated data processing and storage steps, and the imaging hardware. This technique requires only a *single* super-resolved hologram acquired under *wavelength-multiplexed* illumination. As such, the proposed approach achieves a similar performance to that of the state-of-the-art absorbance spectrum estimation method [7] that uses *four* super-resolved holograms collected at *four* sample-to-sensor distances with either sequential or multiplexed illumination wavelengths, thus representing more than four-fold enhancement in terms of data throughput.

We demonstrate the success of this framework using two types of pathology slides: lung tissue sections stained with Masson's trichrome and prostate tissue sections stained with Hematoxylin and Eosin (H&E). Using both the structural similarity index (SSIM) [27] and the color distance [28], high fidelity and color-accurate images are reconstructed and compared to the gold-standard images obtained using the hyperspectral imaging approach. The overall time performance of the proposed framework is also compared against a conventional 20× bright-field scanning microscope, thus demonstrating that the total image acquisition and processing times are of the same scale. We believe that the presented deep learning-based color imaging framework might be helpful to bring coherent microscopy techniques into use for histopathology applications.

## 2. MATERIALS AND METHODS

### 2.1. Overview of the hyperspectral and deep neural network-based reconstruction approaches

We train a deep neural network to perform the image transformation from a complex field obtained from a single super-resolved hologram to the gold-standard image, which is obtained from $N_H \times N_M$ super-resolved holograms ($N_H$ is the number of sample-to-sensor distances, and $N_M$ is the number of measurements at one specific illumination condition). In this work, to generate the gold-standard images using the hyperspectral imaging approach, we used $N_H = 8$ and $N_M = 31$ *sequential* illumination wavelengths (ranging from 400 nm to 700 nm with 10 nm step size). The following subsections detail the procedures used to generate both the gold-standard images as well as the inputs to the deep network.

## 2.2 Hyperspectral imaging approach

The gold-standard, hyperspectral imaging approach reconstructs a high-fidelity color image by first performing resolution enhancement using a PSR algorithm (Section 2.2.1) Subsequently, the missing phase-related artifacts are eliminated using multiheight phase recovery (Section 2.2.3). Finally, high-fidelity color images are generated with tristimulus color projections (Section 2.2.4).

### 2.2.1 Holographic pixel super-resolution using sequential illumination

The resolution enhancement for the hyperspectral imaging approach was performed using a PSR algorithm [12]. This algorithm is capable of digitally synthesizing a high-resolution image (pixel size of approximately 0.37 µm) from a set of low-resolution images collected by an RGB image sensor (IMX 081, Sony, pixel size of 1.12 µm, with R, $G_1$, $G_2$, and B color channels). To acquire these images, the image sensor was programmed to raster through a 6×6 lateral grid using a 3D positioning stage (MAX606, Thorlabs, Inc.) with a subpixel spacing of ~0.37 µm (i.e., 1/3 of the pixel size). At each lateral position, one low-resolution hologram intensity was recorded. The displacement/shift of the sensor was accurately estimated using the algorithm introduced in [14]. A shift-and-add based algorithm was then used to synthesize the high-resolution image.

Because this hyperspectral imaging approach uses *sequential* illumination, the PSR algorithm uses only one color channel (R, $G_1$, or B) from the RGB image sensor at any given illumination wavelength.

Based on the transmission spectral response curves of the Bayer RGB image sensor, the blue channel (B) was used for the illumination wavelengths in the range of 400–470 nm, the green channel ($G_1$) was used for the illumination wavelengths in the range of 480–580 nm, and the red channel (R) was used for the illumination wavelengths in the range of 590–700 nm.

**2.2.2 Angular spectrum propagation**

Free-space angular spectrum propagation [29] was used in the hyperspectral imaging approach to create the ground truth images. To digitally obtain the optical field $U(x,y; z)$ at a propagation distance $z$, the Fourier transform (FT) is first applied to the given $U(x,y; 0)$ to obtain the angular spectrum distribution $A(f_x, f_y; 0)$. The angular spectrum $A(f_x, f_y; z)$ of the optical field $U(x,y; z)$ can be calculated using:

$$A(f_x, f_y; z) = A(f_x, f_y; 0) \cdot H(f_x, f_y; z) \quad (1)$$

where $H(f_x, f_y; z)$ is defined as,

$$H(f_x, f_y; z) = \begin{cases} 0, & \left(\dfrac{\lambda f_x}{n}\right)^2 + \left(\dfrac{\lambda f_y}{n}\right)^2 > 1 \\ \exp\left[j2\pi \dfrac{n}{\lambda} z \sqrt{1 - \left(\dfrac{\lambda f_x}{n}\right)^2 - \left(\dfrac{\lambda f_y}{n}\right)^2}\right], & \text{otherwise} \end{cases} \quad (2)$$

where $\lambda$ is the illumination wavelength, and $n$ is the refractive index of the medium. Finally, an inverse Fourier transform is applied to $A(f_x, f_y; z)$ to get $U(x,y; z)$.

This angular spectrum propagation method first served as the building block of an autofocusing algorithm, which is used to estimate the sample to sensor distance for each acquired hologram [30,31]. After the accurate sample to the sensor distances were estimated, the hyperspectral imaging approach used the angular spectrum propagation as an additional building block for the iterative multiheight phase recovery, which will be detailed next.

**2.2.3 Multiheight phase recovery**

To eliminate the spatial image artifacts related to the missing phase, the hyperspectral imaging approach applied an iterative phase retrieval algorithm [13]. Holograms from eight sample-to-sensor distances were collected during the data acquisition step. The algorithm initially assigned a zero-phase to the intensity measurement of the object. Each iteration of the algorithm began by propagating the complex field from the first height to the eighth height, and by backpropagating it to the first height. The amplitude was updated at each height, while the phase was kept unchanged. The algorithm typically converged after 10–30 iterations. Finally, the complex field was backpropagated from any one of the measurement planes to the object plane to retrieve both the amplitude and the phase images.

### 2.2.4 Color tristimulus projection

Increased color accuracy was achieved by densely sampling the visible band at 31 different wavelengths in the range of 400 nm to 700 nm at a 10 nm step size. This spectral information was projected to a color tristimulus using the Commission Internationale de l'Éclairage (CIE) color matching function [6]. The color tristimulus in the XYZ color space can be calculated by,

$$\begin{aligned} X &= \int \bar{x}(\lambda) T(\lambda) E(\lambda) \mathrm{d}\lambda \\ Y &= \int \bar{y}(\lambda) T(\lambda) E(\lambda) \mathrm{d}\lambda \\ Z &= \int \bar{z}(\lambda) T(\lambda) E(\lambda) \mathrm{d}\lambda \end{aligned} \quad (3)$$

where $\lambda$ is the wavelength, $\bar{x}(\lambda)$, $\bar{y}(\lambda)$, and $\bar{z}(\lambda)$ are the CIE color matching functions, $T(\lambda)$ is the transmittance spectrum of the sample, and $E(\lambda)$ is the CIE standard illuminant D65 [6]. The XYZ values can be linearly transformed to the standard RGB values for display [6].

### 2.3 High-fidelity holographic color reconstruction via deep neural networks

The input complex fields for the proposed deep learning-based color reconstruction framework were generated in the following manner: Resolution enhancement and cross-talk correction through the demosaiced pixel super resolution algorithm (Section 2.3.1) followed by the initial estimation of the object via the angular spectrum propagation (Section 2.2.2).

**2.3.1 Holographic demosaiced pixel super-resolution (DPSR) using multiplexed illumination**

Similar to the hyperspectral imaging approach, the proposed network approach also used a shift-and-add-based algorithm in association with 6×6 low-resolution holograms to enhance the hologram resolution. We used three multiplexed wavelengths, i.e., simultaneously illuminated the sample with three distinct wavelengths. To correct the cross-talk error among different color channels in the RGB sensor we used the DPSR algorithm [26]. This cross-talk correction can be illustrated by the following equation:

$$\begin{bmatrix} U_R \\ U_G \\ U_B \end{bmatrix} = \mathbf{W} \times \begin{bmatrix} U_{R\_ori} \\ U_{G_1\_ori} \\ U_{G_2\_ori} \\ U_{B\_ori} \end{bmatrix} \quad (4)$$

where $U_{R\text{-ori}}$, $U_{G_1\text{-ori}}$, $U_{G_2\text{-ori}}$, and $U_{B\text{-ori}}$, represent the original interference patterns collected by the image sensor, $\mathbf{W}$ is a 3×4 cross-talk matrix obtained by experimental calibration of a given RGB sensor chip, and $U_R$, $U_G$, and $U_B$, are the demultiplexed (R, G, B) interference patterns. In this work, the three illumination wavelengths were chosen to be at 450 nm, 540 nm, and 590 nm. As suggested in [7], using these wavelengths, a better color accuracy can be achieved with specific tissue-stain types (i.e., prostate stained with H&E and lung stained with Masson's trichrome, which were used in this work).

**2.3.2 Deep neural network input formation**

Following the demosaiced pixel-super-resolution algorithm, the three intensity holograms are numerically backpropagated to the object plane, as discussed in Subsection 2.2.2. Following this backpropagation step, each one of the three color hologram channels will produce a complex wave, represented as real and imaginary data channels. This results in a six-channel tensor that is used as input to the deep network, as shown in **Figure 1**. Unlike the ground truth, in this case, no phase retrieval is performed because only a single measurement is available.

**2.3.3 Deep neural network architecture**

A generative adversarial network (GAN [32]) was implemented to learn the color correction and eliminate the missing phase-related artifacts. This GAN framework has recently found applications in super-resolution microscopic imaging [33–35] and histopathology [36,37], and it consists of a discriminator network ($D$) and a generator network ($G$). The $D$ network was used to distinguish between a three-channel RGB ground truth image ($z$) obtained from hyperspectral imaging and the output image from $G$. Accordingly, $G$ was used to learn the transformation from a six-channel holographic image ($x$), i.e., three color channels with real and imaginary components, into the corresponding RGB ground truth image.

Our discriminator and generator losses are defined as,

$$l_{\text{discriminator}} = D(G(x))^2 + (1 - D(z))^2 \tag{5}$$

$$l_{\text{generator}} = L_2\{z, G(x)\} + \lambda \times TV\{G(x)\} + \alpha \times (1 - D(G(x)))^2 \tag{6}$$

where,

$$L_2\{z, G(x)\} = \frac{1}{N_{\text{channels}} \times M \times N} \sum_{n=1}^{N_{\text{channels}}} \sum_{i,j=1}^{M,N} (x_{i,j,n} - z_{i,j,n})^2 \tag{7}$$

where $N_{\text{channels}}$ is the number of channels in the images (e.g., $N_{\text{channels}} = 3$ for an RGB image), $M$ and $N$ are the number of pixels for each side of the images, $i$ and $j$ are the pixel indices, and $n$ denotes the channel indices. $TV$ represents the total variation regularizer that applies to the generator output, and is defined as,

$$TV(x) = \frac{1}{N_{\text{channels}}} \sum_{n=1}^{N_{\text{channels}}} \sum_{i,j=1}^{M,N} \left( \left| x_{i+1,j,n} - x_{i,j,n} \right| + \left| x_{i,j+1,n} - x_{i,j,n} \right| \right) \tag{8}$$

The regularization parameters ($\lambda$, $\alpha$) were set to 0.0025 and 0.002 so that the total variation loss ($\lambda \times TV\{G(x_{\text{input}})\}$) is ~2% of $L_2$, and the discriminator loss ($\alpha \times (1-D(G(x_{\text{input}})))^2$) is ~15% of $l_{\text{generator}}$. Ideally, both $D(z_{\text{label}})$ and $D(G(x_{\text{input}}))$ converge to 0.5 at the end of the training phase. The L2-loss was empirically found to better handle distortions, which resulted due to the free-space back propagation of the single RGB DPSR hologram.

The generator network architecture (**Figure 2**) was an adapted form of the U-net [38]. Additionally, the discriminator network (see **Figure 3**) used a simple classifier that consisted of a series of convolutional layers which slowly reduced the dimensionality, while they increased the number of channels, followed by two fully connected layers to output the classification. While in this manuscript we adapted the U-net structure for our deep network, other structures can also be considered for elimination of missing phase artifacts [39] and for performing color correction on the reconstructed images. The convolution filter size was set to 3×3, and each convolutional layer except the last was followed by a leaky-ReLu activation function, defined as:

$$leaky\text{-}ReLu(x) = \begin{cases} x & \text{for } x > 0 \\ 0.1x & \text{otherwise} \end{cases} \quad (9)$$

**2.3.4 Deep neural network training process**

In the network training process, we used the images generated by the hyperspectral approach as our network labels, and took the demosaiced super-resolved holograms that were back-propagated to the sample plane as the network inputs. Both the generator and the discriminator networks were trained with non-overlapping patches, each with a size of 128×128 pixels. The weights in the convolutional layers and fully connected layers, were initialized using the Xavier initialization [40] while the biases were initialized to zero. All parameters were updated using an adaptive moment estimation (Adam) optimizer [41] with a learning rate of $1\times10^{-4}$ for the generator network and a corresponding rate of $5\times10^{-5}$ for the discriminator network. The training, validation, and testing of the network were performed on a PC with a four-core 3.60 GHz CPU, 16 GB of RAM, and an Nvidia GeForce GTX 1080Ti GPU. The lung tissue network was trained for 38.9 epochs over 5.58 hours, while the prostate tissue network was trained for 25.6 epochs over 2.29 hours. This training process only needs to be performed once for a specific type of tissue-stain combination and can improve in speed by using cloud computing.

**2.4. Bright-field imaging**

For comparison of the imaging throughput, bright-field microscopy images were obtained. An Olympus IX83 microscope equipped with a motorized stage and a set of super panchromatic objectives (Olympus UPLSAPO 20×/0.75 numerical aperture (NA), working distance (WD) 0.65) were used. The microscope was controlled by the MetaMorph advanced digital imaging software (Version 7.10.1.161, MetaMorph®) with the autofocusing algorithm set to search in a range of 5 μm in the z–direction with 1 μm accuracy. Two-pixel binning was enabled and a 10% overlap between the scanned patches was used. Stitching was done using the ImageJ Grid/Collection stitching plugin [42], which calculates the exact overlap between the images, and linearly blends the overlapping section, which allows the image to have a smooth transition and reduces stitching related artifacts.

**2.5. Quantification metrics**

Quantification metrics were chosen and used to evaluate the performance of the network: the SSIM [27] was used to compare the similarity of the tissue structural information between the output and the target images; ΔE*94 [28] was used to compare the color distance of the two images.

SSIM values ranged from zero to one, whereby the value of unity indicated that the two images were the same, i.e.,

$$\text{SSIM}(U,V) = \frac{(2\mu_U \mu_V + C_1)(2\sigma_{U,V} + C_2)}{(\mu_U^2 + \mu_V^2 + C_1)(\sigma_U^2 + \sigma_V^2 + C_2)} \tag{10}$$

where $U$ and $V$ represent one vectorized test image and one vectorized reference image, respectively, $\mu_U$ and $\mu_V$ are the means of $U$ and $V$, respectively, $\sigma_U^2, \sigma_V^2$ are the variances of $U$ and $V$, respectively, $\sigma_{U,V}$ is the covariance of $U$ and $V$, and constants $C_1$ and $C_2$ are included to stabilize the division when the denominator is close to zero.

The second metric that we used, ΔE*94 [28], outputs a number between zero and 100. A value of zero indicates that the compared pixels share the exact same color, while a value of 100 indicates that the two images have the opposite color (mixing two opposite colors cancel each other out and produce a gray-scale color). This method calculates the color distance in a pixel-wise fashion, and the final result is calculated by averaging the values of ΔE*94 in every pixel of the output image.

**2.6. Sample preparation**

De-identified H&E stained human prostate tissue slides and Masson's trichrome stained human lung tissue slides were acquired from the UCLA Translational Pathology Core Laboratory. Existing and anonymous specimens were used. No subject related information was linked or can be retrieved.

**3. RESULTS AND DISCUSSION**

**3.1. Qualitative assessment**

We evaluated our network's performance using two different tissue-stain combinations: prostate tissue sections stained with H&E, and lung tissue sections stained with Masson's trichrome. For both types of samples, the networks were trained on three tissue sections from different patients and were blindly tested on another tissue section from a fourth patient. The field-of-view (FOV) of each tissue section that was used for training and testing was ~20 mm$^2$.

The results for lung and prostate samples are respectively summarized in **Figures 4** and **5**. These indicate our approach's capability of reconstructing a high-fidelity and color-accurate image from a single nonphase-retrieved and wavelength-multiplexed hologram (as detailed in the Methods section). Using the trained model, we were able to reconstruct the sample image over the entire sensor's FOV (i.e., ~20 mm$^2$), as demonstrated in **Figure 6**.

To further demonstrate the qualitative performance of the network, we compare in **Figures 7** and **8** the reconstruction results of the deep network to the images created by the absorbance spectrum estimation method [7] in terms of the required number of measurements. For this comparison, we implemented the spectrum estimation approach for the multiheight phase recovery method and reconstructed the color images from a reduced number of wavelengths via both sequential ($N_H$=8, $N_M$=3) and multiplexed ($N_H$=8,

$N_M$=1) illuminations at the same wavelengths (i.e., 450 nm, 540 nm, and 590 nm). Qualitatively, the network results are comparable to the multiheight results obtained with more than four sample-to-sensor distances for both the sequential and multiplexed illumination cases. This will be also confirmed by the quantitative analysis described below.

### 3.2. Quantitative performance assessment

The quantitative performance of the network was evaluated based on the calculation of the SSIM [27] and color difference (ΔE*94 [28]) between the network's output and the gold-standard image produced by the hyperspectral imaging approach. As listed in Table 1 and visually shown in **Figures 5 and 6**, the performances of the spectrum estimation methods decrease (i.e., SSIM decreases and ΔE*94 increases) as the number of holograms at different sample-to-sensor distances decreases, or when the illumination is changed to be multiplexed. This quantitative comparison demonstrates that the network's performance using a single super-resolved hologram is comparable to the results obtained by state-of-the-art algorithms where ≥4 times as many raw holographic measurements are used.

### 3.3. Throughput evaluation

Table 2 lists the measured reconstruction times for the entire FOV (~20 mm$^2$) using different methods. For the deep neural network approach, the total reconstruction time includes the acquisition of 36 holograms (at 6×6 lateral positions in multiplexed illumination), the execution of DPSR, angular spectrum propagation, network inference, and image stitching. For the hyperspectral imaging approach, the total reconstruction time includes the collection of 8928 holograms (at 6×6 lateral positions, eight sample-to-sensor distances, and 31 wavelengths), PSR, multiheight phase retrieval, color tristimulus projection, and image stitching. For the conventional bright-field microscope (equipped with an automatic scanning stage), the total time includes the scanning of the bright-field images using a 20×/0.75 NA microscope with autofocusing performed at each scanning position and image stitching. In addition, the timing of the multiheight phase recovery method with the use of four sample-to-sensor distances was also shown, and had the closest performance to the deep learning-based neural network approach. All the coherent imaging related algorithms were accelerated with an Nvidia GTX 1080Ti GPU and CUDA C++ programming.

The network-based method took ~7 min to acquire and reconstruct a 20 mm$^2$ tissue area, which was approximately equal to the time it would take to image the same region using the 20× objective with our standard, general-purpose, bright-field scanning microscope. This is significantly shorter than the ~60 min required when using the spectral estimation approach (with four heights and simultaneous illumination). The deep learning approach also increases the data efficiency. The raw super-resolved hologram data size was reduced from 4.36 GB to 1.09 GB, which is more comparable to the data size of bright-field scanning microscopy images, which in total used 577.13 MB.

## 4. CONCLUSIONS

We presented a deep learning-based color holographic imaging system and demonstrated its performance using histologically stained pathology slides. This framework significantly simplified the data acquisition procedure, reduced the data storage requirement, shortened the processing time, and enhanced the color accuracy of the holographically reconstructed images. It is important to note that other technologies, such as slide-scanner microscopes used in pathology can readily scan tissue slides at much faster rates, although they are rather expensive for use in resource limited settings. Therefore, further improvements to our lensless holographic imaging hardware, such as for example, the use of illumination arrays to perform pixel super resolution [19] would be needed to improve the overall reconstruction time of our results.

**Acknowledgments**

The Ozcan Research Group at UCLA acknowledges the support of NSF Engineering Research Center (ERC, PATHS-UP).

# REFERENCES

1. W. C. Revie, M. Shires, P. Jackson, D. Brettle, R. Cochrane, and D. Treanor, "Color management in digital pathology," Analytical Cellular Pathology **2014**, (2014).
2. W. S. Campbell, G. A. Talmon, K. W. Foster, S. M. Lele, J. A. Kozel, and W. W. West, "Sixty-five thousand shades of gray: importance of color in surgical pathology diagnoses," Human pathology **46**, 1945–1950 (2015).
3. P. A. Bautista, N. Hashimoto, and Y. Yagi, "Color standardization in whole slide imaging using a color calibration slide," Journal of pathology informatics **5**, (2014).
4. P. Shrestha and B. Hulsken, "Color accuracy and reproducibility in whole slide imaging scanners," Journal of Medical Imaging **1**, 027501 (2014).
5. M. S. Peercy and L. Hesselink, "Wavelength selection for true-color holography," Applied optics **33**, 6811–6817 (1994).
6. D. Pascale, "A review of rgb color spaces... from xyy to r'g'b'," Babel Color **18**, 136–152 (2003).
7. Y. Zhang, T. Liu, Y. Huang, D. Teng, Y. Bian, Y. Wu, Y. Rivenson, A. Feizi, and A. Ozcan, "Accurate color imaging of pathology slides using holography and absorbance spectrum estimation of histochemical stains," Journal of Biophotonics e201800335 (2018).
8. S. G. Kalenkov, G. S. Kalenkov, and A. E. Shtanko, "Hyperspectral digital holography of microobjects," in *Practical Holography XXIX: Materials and Applications* (International Society for Optics and Photonics, 2015), Vol. 9386, p. 938604.
9. P. Xia, Y. Ito, Y. Shimozato, T. Tahara, T. Kakue, Y. Awatsuji, K. Nishio, S. Ura, T. Kubota, and O. Matoba, "Digital Holography Using Spectral Estimation Technique," J. Display Technol., JDT **10**, 235–242 (2014).
10. A. Greenbaum, W. Luo, T.-W. Su, Z. Göröcs, L. Xue, S. O. Isikman, A. F. Coskun, O. Mudanyali, and A. Ozcan, "Imaging without lenses: achievements and remaining challenges of wide-field on-chip microscopy," Nat. Methods **9**, 889–895 (2012).
11. L. J. Allen and M. P. Oxley, "Phase retrieval from series of images obtained by defocus variation," Optics Communications **199**, 65–75 (2001).
12. A. Greenbaum, Y. Zhang, A. Feizi, P.-L. Chung, W. Luo, S. R. Kandukuri, and A. Ozcan, "Wide-field computational imaging of pathology slides using lens-free on-chip microscopy," Science Translational Medicine **6**, 267ra175-267ra175 (2014).
13. A. Greenbaum and A. Ozcan, "Maskless imaging of dense samples using pixel super-resolution based multi-height lensfree on-chip microscopy," Opt. Express, OE **20**, 3129–3143 (2012).
14. A. Greenbaum, U. Sikora, and A. Ozcan, "Field-portable wide-field microscopy of dense samples using multi-height pixel super-resolution based lensfree imaging," Lab Chip **12**, 1242–1245 (2012).
15. Y. Rivenson, Y. Wu, H. Wang, Y. Zhang, A. Feizi, and A. Ozcan, "Sparsity-based multi-height phase recovery in holographic microscopy," Scientific Reports **6**, 37862 (2016).
16. W. Luo, A. Greenbaum, Y. Zhang, and A. Ozcan, "Synthetic aperture-based on-chip microscopy," Light: Science & Applications **4**, e261 (2015).
17. W. Luo, Y. Zhang, A. Feizi, Z. Göröcs, and A. Ozcan, "Pixel super-resolution using wavelength scanning," Light Sci Appl. **5**, e16060 (2016).
18. W. Bishara, T.-W. Su, A. F. Coskun, and A. Ozcan, "Lensfree on-chip microscopy over a wide field-of-view using pixel super-resolution," Optics Express **18**, 11181 (2010).
19. W. Bishara, U. Sikora, O. Mudanyali, T.-W. Su, O. Yaglidere, S. Luckhart, and A. Ozcan, "Holographic pixel super-resolution in portable lensless on-chip microscopy using a fiber-optic array," Lab on a Chip **11**, 1276 (2011).
20. A. Greenbaum and A. Ozcan, "Maskless imaging of dense samples using pixel super-resolution based multi-height lensfree on-chip microscopy," Optics Express **20**, 3129 (2012).

21. A. Greenbaum, U. Sikora, and A. Ozcan, "Field-portable wide-field microscopy of dense samples using multi-height pixel super-resolution based lensfree imaging," Lab on a Chip **12**, 1242 (2012).
22. S. O. Isikman, W. Bishara, and A. Ozcan, "Lensfree On-chip Tomographic Microscopy Employing Multi-angle Illumination and Pixel Super-resolution," Journal of Visualized Experiments (2012).
23. A. Greenbaum, N. Akbari, A. Feizi, W. Luo, and A. Ozcan, "Field-Portable Pixel Super-Resolution Colour Microscope," PLoS ONE **8**, e76475 (2013).
24. A. Greenbaum, A. Feizi, N. Akbari, and A. Ozcan, "Wide-field computational color imaging using pixel super-resolved on-chip microscopy," Optics Express **21**, 12469 (2013).
25. A. Greenbaum, W. Luo, B. Khademhosseinieh, T.-W. Su, A. F. Coskun, and A. Ozcan, "Increased space-bandwidth product in pixel super-resolved lensfree on-chip microscopy," Sci. Rep. **3**, (2013).
26. Y. Wu, Y. Zhang, W. Luo, and A. Ozcan, "Demosaiced pixel super-resolution for multiplexed holographic color imaging," Sci Rep **6**, (2016).
27. Z. Wang, A. C. Bovik, H. R. Sheikh, and E. P. Simoncelli, "Image quality assessment: from error visibility to structural similarity," IEEE transactions on image processing **13**, 600–612 (2004).
28. B. Hill, T. Roger, and F. W. Vorhagen, "Comparative analysis of the quantization of color spaces on the basis of the CIELAB color-difference formula," ACM Transactions on Graphics (TOG) **16**, 109–154 (1997).
29. J. W. Goodman, *Introduction to Fourier Optics* (Roberts and Company Publishers, 2005).
30. Y. Zhang, H. Wang, Y. Wu, M. Tamamitsu, and A. Ozcan, "Edge sparsity criterion for robust holographic autofocusing," Optics Letters **42**, 3824 (2017).
31. M. Tamamitsu, Y. Zhang, H. Wang, Y. Wu, and A. Ozcan, "Comparison of Gini index and Tamura coefficient for holographic autofocusing based on the edge sparsity of the complex optical wavefront," arXiv:1708.08055 [physics.optics] (2017).
32. I. Goodfellow, J. Pouget-Abadie, M. Mirza, B. Xu, D. Warde-Farley, S. Ozair, A. Courville, and Y. Bengio, "Generative adversarial nets," in *Advances in Neural Information Processing Systems* (2014), pp. 2672–2680.
33. H. Wang, Y. Rivenson, Y. Jin, Z. Wei, R. Gao, H. Günaydın, L. A. Bentolila, C. Kural, and A. Ozcan, "Deep learning enables cross-modality super-resolution in fluorescence microscopy," Nature Methods **16**, 103–110 (2019).
34. T. Liu, K. de Haan, Y. Rivenson, Z. Wei, X. Zeng, Y. Zhang, and A. Ozcan, "Deep learning-based super-resolution in coherent imaging systems," Scientific reports **9**, 3926 (2019).
35. K. de Haan, Z. S. Ballard, Y. Rivenson, Y. Wu, and A. Ozcan, "Resolution enhancement in scanning electron microscopy using deep learning," arXiv:1901.11094 [physics] (2019).
36. Y. Wu, Y. Luo, G. Chaudhari, Y. Rivenson, A. Calis, K. de Haan, and A. Ozcan, "Bright-field holography: cross-modality deep learning enables snapshot 3D imaging with bright-field contrast using a single hologram," Light: Science & Applications **8**, 25 (2019).
37. Y. Rivenson, H. Wang, Z. Wei, K. de Haan, Y. Zhang, Y. Wu, H. Günaydın, J. E. Zuckerman, T. Chong, and A. E. Sisk, "Virtual histological staining of unlabelled tissue-autofluorescence images via deep learning," Nature Biomedical Engineering 1 (2019).
38. O. Ronneberger, P. Fischer, and T. Brox, "U-Net: Convolutional Networks for Biomedical Image Segmentation," arXiv:1505.04597 [cs] (2015).
39. Y. Rivenson, Y. Zhang, H. Günaydın, D. Teng, and A. Ozcan, "Phase recovery and holographic image reconstruction using deep learning in neural networks," Light: Science & Applications **7**, 17141 (2018).
40. X. Glorot and Y. Bengio, "Understanding the difficulty of training deep feedforward neural networks. 2010," in *Internaional Conference on Artificial Intelligence and Statistics* (n.d.).
41. D. P. Kingma and J. Ba, "Adam: A method for stochastic optimization," arXiv preprint arXiv:1412.6980 (2014).

42. S. Preibisch, S. Saalfeld, and P. Tomancak, "Globally optimal stitching of tiled 3D microscopic image acquisitions," Bioinformatics **25**, 1463–1465 (2009).

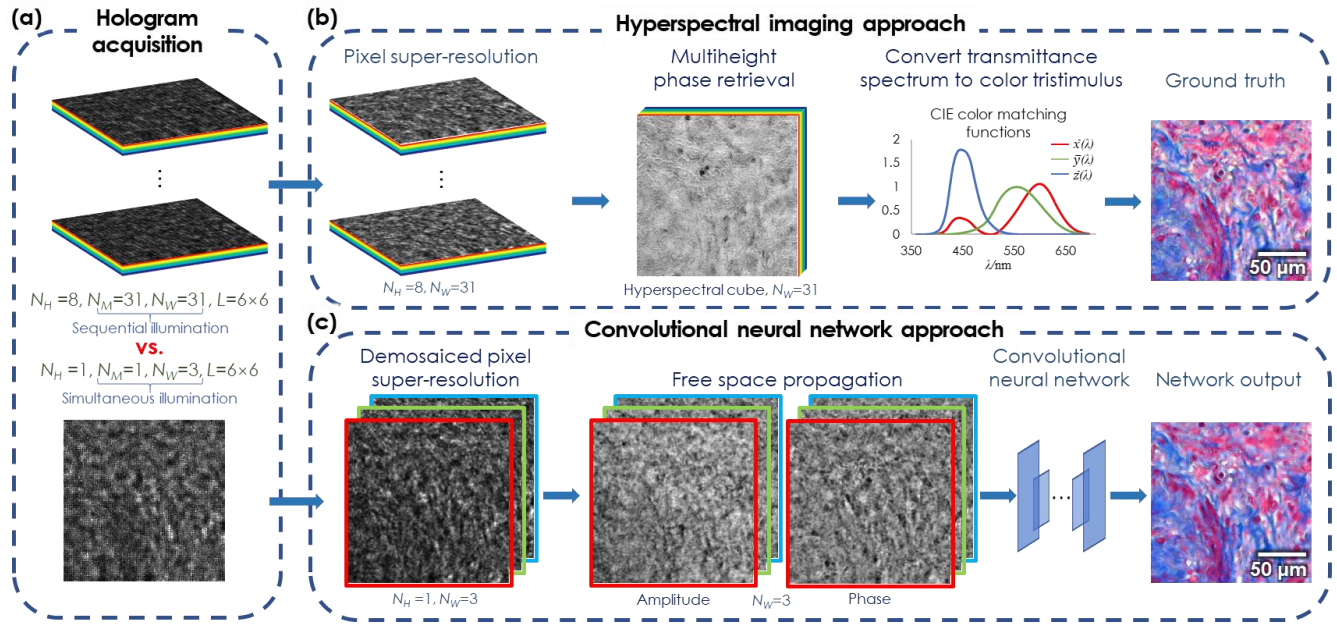

**Figure 1.** Comparison between the traditional hyperspectral imaging and the proposed neural network-based approaches for the reconstruction of accurate color images. $N_H$ is the number of sample-to-sensor heights required for performing phase recovery, $N_W$ is the number of illumination wavelengths, $N_M$ is the number of measurements for each illumination condition (multiplexed or sequential), and $L$ is the number of lateral positions used to perform pixel super resolution. **(a)**: Required number of raw holograms for the traditional hyperspectral imaging and the proposed neural network-based approaches. **(b)**: High fidelity color image reconstruction procedure for the hyperspectral imaging approach. **(c)**: High fidelity color image reconstruction procedure for the proposed neural network-based approach.

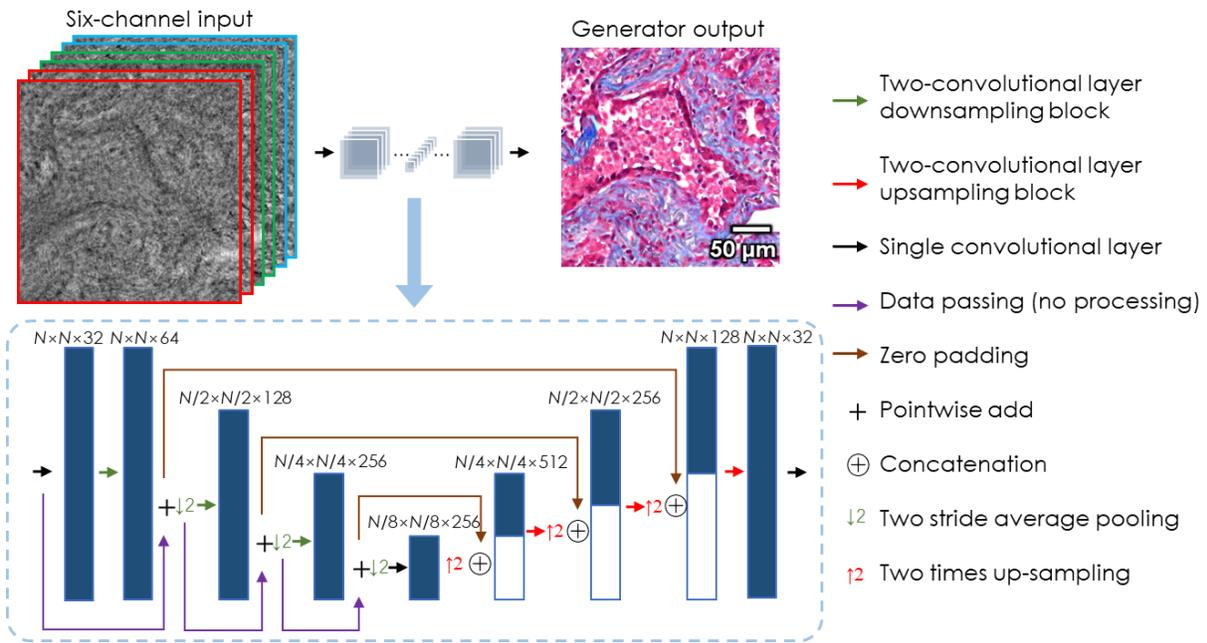

**Figure 2.** Schematic of the generator part of the network. The six-channel input consists of the real and imaginary channels of the three free-space propagated holograms at three illumination wavelengths (450 nm, 540 nm, and 590 nm). Each down block consists of two convolutional layers that double the number of system channels when used together. The down blocks are opposite, and consist of two convolutional layers with half the number of system channels when used together.

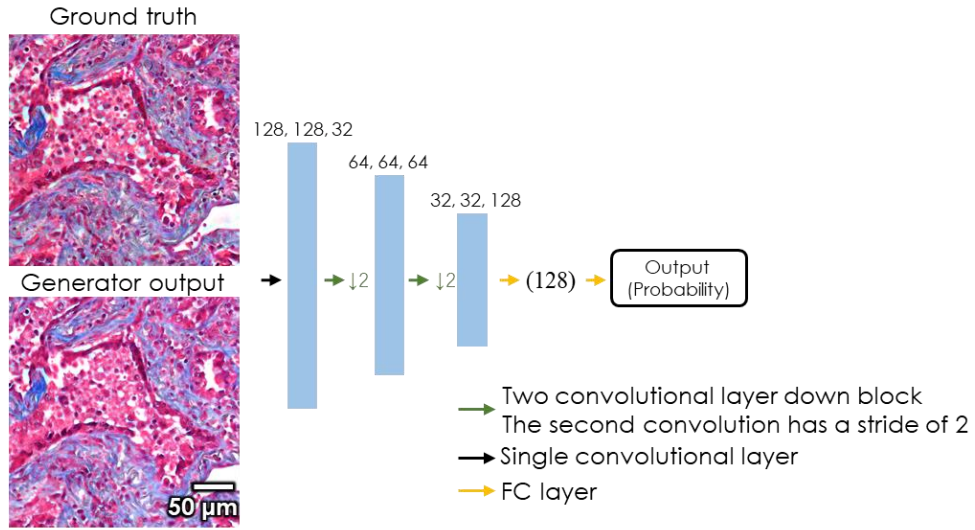

**Figure 3.** Diagram of the discriminator part of the network. Each down block of the convolutional layer consists of two convolutional layers.

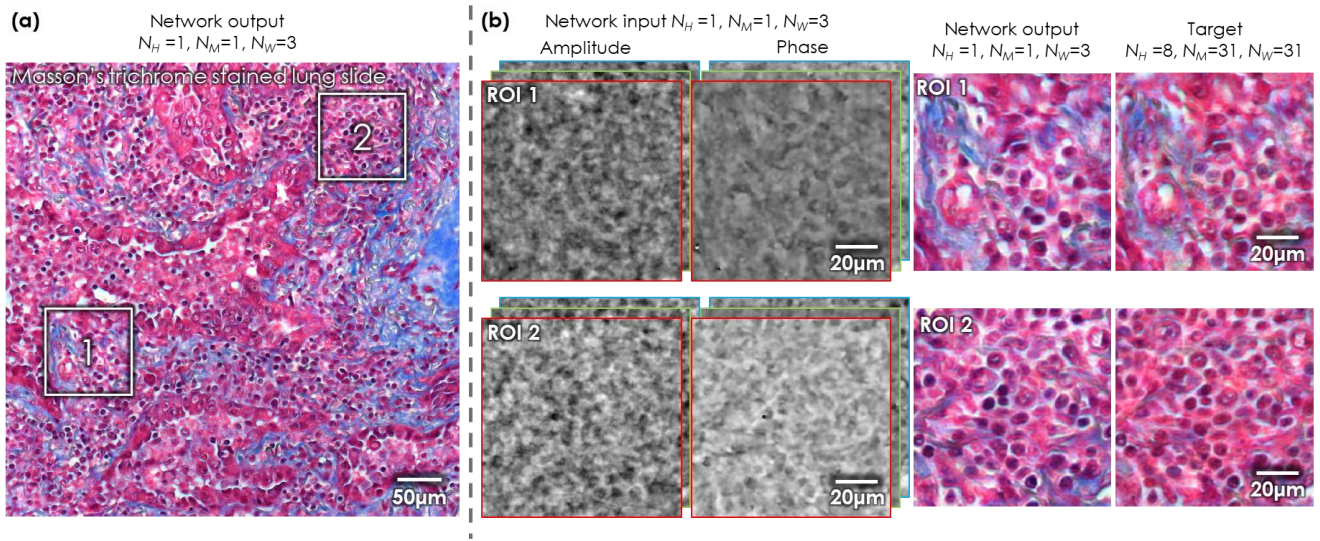

**Figure 4.** Deep learning-based accurate color imaging of a lung tissue slide stained with Masson's trichrome for a multiplexed illumination at 450 nm, 540 nm, and 590 nm, using a lens-free holographic on-chip microscope. **(a)**: Large field of view of the network output image. **(b)**: Zoomed-in comparison of the network input, the network output, and the ground truth target at region of interest (ROI) 1 and 2.

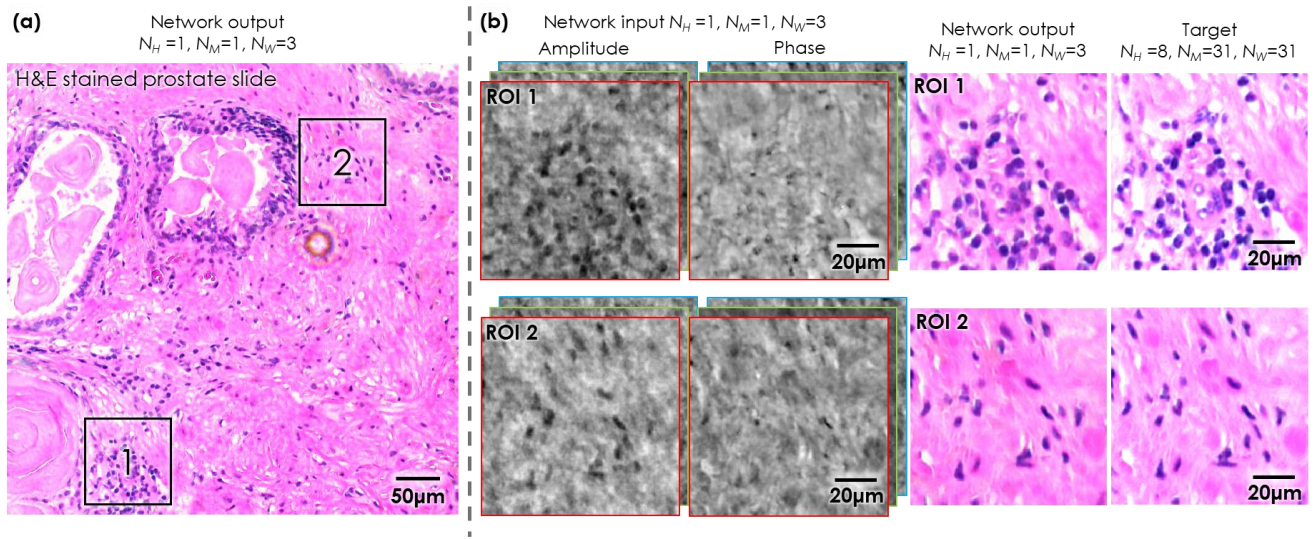

**Figure 5.** Deep learning-based accurate color imaging of a prostate tissue slide stained with H&E for a multiplexed illumination at 450 nm, 540 nm, and 590 nm, using a lens-free holographic on-chip microscope. **(a)**: Large field of view of the network output image. **(b)** Zoomed-in comparison of the network input, the network output, and the ground truth target at ROI 1 and 2.

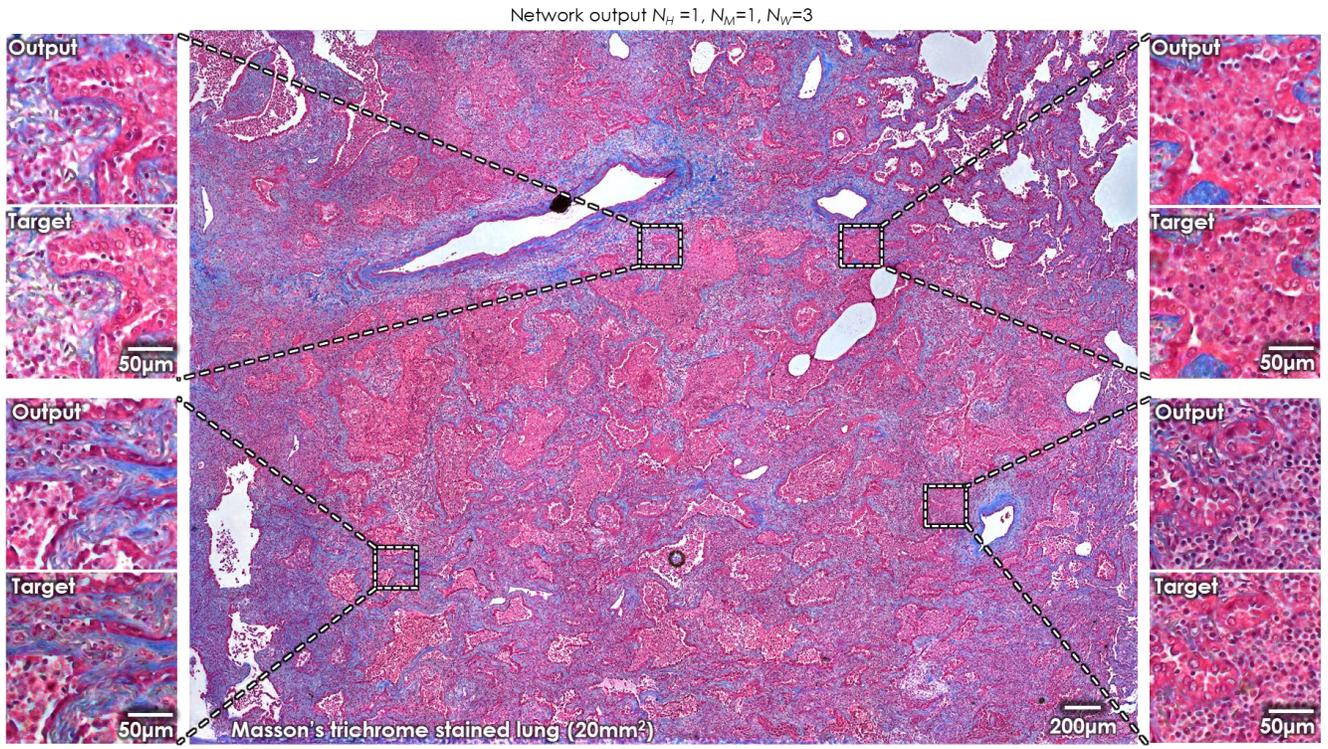

**Figure 6.** Stitched image of the deep neural network output for a lung tissue section stained with H&E, which corresponds to the sensor's field-of-view.

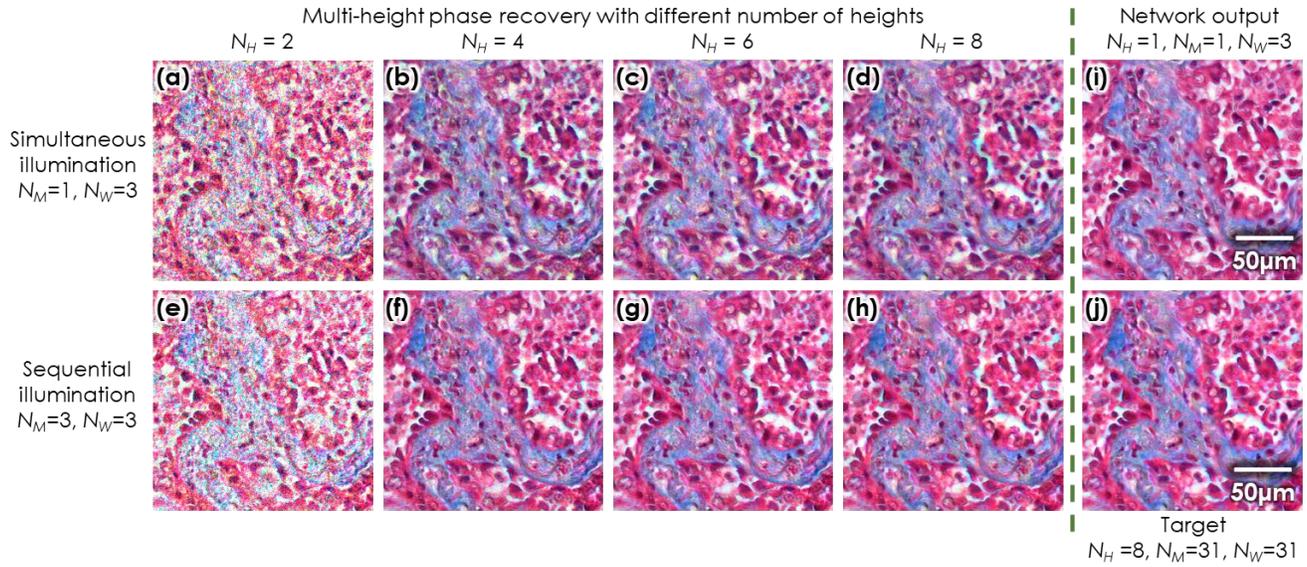

**Figure 7.** Visual comparison between the deep neural network-based approach and the multiheight phase recovery with spectral estimation approach for a lung tissue sample stained with Masson's trichrome. **(a-h)**: Reconstruction results of spectral estimation approach using different number of heights and different illumination conditions. **(i)**: Network output. **(j)**: Ground truth target obtained using the hyper-spectral imaging approach.

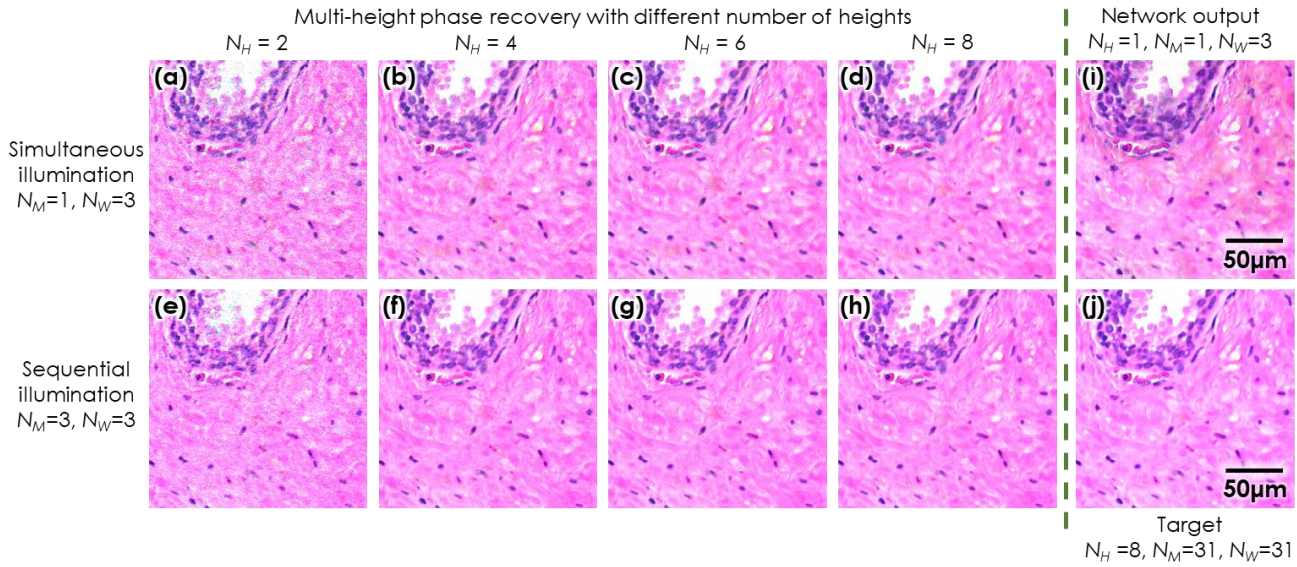

**Figure 8.** Visual comparison between the deep neural network-based approach and the multiheight phase recovery with the spectral estimation approach for a prostate tissue sample stained with H&E. **(a-h)**: Reconstruction results of spectral estimation approach using different number of heights and different illumination conditions. **(i)**: Network output. **(j)**: Ground truth target obtained using the hyperspectral imaging approach.

**Table 1. Comparison of SSIM) and ΔE*94 performances between the deep neural network approach and various other methods using two, four, six, and eight sample-to-sensor heights and three sequential/multiplexed wavelength illumination conditions for two tissue samples (the network-based approach and other methods with comparable performance are highlighted with bold font).**

| Tissue-stain type | Method | Illumination condition (at 450 nm, 540 nm, and 590 nm) | Total required measurements ($N_H \times N_M \times L$) | Average SSIM | ΔE*94 |
|---|---|---|---|---|---|
| Masson's tri-chrome stained lung slide (~20 mm² FOV) | Deep neural network | Simultaneous | 1×1×36 | **0.8396** | 6.9044 |
| | Two-height reconstruction | Simultaneous | 2×1×36 | 0.5535 | 10.7507 |
| | | Sequential | 2×3×36 | 0.6011 | 9.4786 |
| | Four-height reconstruction | Simultaneous | 4×1×36 | **0.8344** | 5.1674 |
| | | Sequential | 4×3×36 | **0.8769** | 3.8709 |
| | Six-height reconstruction | Simultaneous | 6×1×36 | **0.878** | 4.4219 |
| | | Sequential | 6×3×36 | 0.9136 | 3.1928 |
| | Eight-height reconstruction | Simultaneous | 8×1×36 | 0.9068 | 3.6779 |
| | | Sequential | 8×3×36 | 0.9538 | 2.1849 |
| Hematoxylin and Eosin stained prostate slide (~20 mm² FOV) | Deep neural network | Simultaneous | 1×1×36 | **0.9249** | 4.5228 |
| | Two-height reconstruction | Simultaneous | 2×1×36 | 0.7716 | 7.5085 |
| | | Sequential | 2×3×36 | 0.848 | 5.5316 |
| | Four-height reconstruction | Simultaneous | 4×1×36 | **0.8984** | 4.3878 |
| | | Sequential | 4×3×36 | **0.9335** | 3.3399 |
| | Six-height reconstruction | Simultaneous | 6×1×36 | **0.9225** | 3.8911 |
| | | Sequential | 6×3×36 | 0.9516 | 2.9622 |
| | Eight-height reconstruction | Simultaneous | 8×1×36 | 0.9411 | 3.5102 |
| | | Sequential | 8×3×36 | 0.9689 | 2.4148 |

**Table 2. Time performance evaluation of the deep neural network approach for reconstructing accurate color images compared to traditional hyperspectral imaging approach and standard brightfield microscopic sample scanning (where N/A stands for "not applicable").**

| Testing area | Method | Data acquisition time | Processing time | | | | | Total time | Storage space (raw data) |
|---|---|---|---|---|---|---|---|---|---|
| | | | Auto-Focusing | Super resolution | Phase recovery or FSP | Inference or color transformation | Stitching | | |
| Sensor's entire FOV ~20 mm² | Deep neural network | ~2 min | ~ 20 s | ~2 min | ~ 3 s | ~ 1.5 min | ~1 min | ~7 min | 1.09 GB |
| | Four-height simultaneous | ~ 8 min | ~ 80 s | ~ 9 min | ~ 5 min | ~36 min | ~1 min | ~60 min | 4.36 GB |
| | Four-height sequential | ~ 25 min | ~ 80 s | ~ 9 min | ~ 5 min | ~36 min | ~1 min | ~77 min | 13.08 GB |
| | Hyperspectral imaging | ~ 8 h | ~ 27 min | ~ 3 h | ~ 85 min | ~15 min | ~1 min | ~13 h | 270.32 GB |
| | Conventional microscope (20×/0.75 NA) | ~6 min | N/A | N/A | N/A | N/A | ~1 min | ~7 min | 577.13 MB |